\documentclass[%showpacs, showkeys,
12pt,
preprint,preprintnumbers,nofootinbib,
groupedaddress,superscriptaddress,amsmath,amssymb]{revtex4}
\usepackage{graphicx}% Include figure files
\usepackage{dcolumn}% Align table columns on decimal point
\usepackage{bm}% bold math
\usepackage{amssymb}
\usepackage{amsmath}
\usepackage{epsfig}    
\usepackage{color}
\usepackage{slashed}
\usepackage{hhline}
\def\be{\begin{equation}}
\def\ee{\end{equation}}
\newcommand{\bea}{\begin{eqnarray}}
\newcommand{\eea}{\end{eqnarray}}
\newcommand{\nn}{\nonumber}

\numberwithin{equation}{section}

\begin{document}

{\begin{flushright}{KIAS-P17055}
\end{flushright}}

%%%%%%%%%
\title{Neutrino mass in flavor dependent gauged lepton model}
%\preprint{KIAS-P14078}
%

\author{Takaaki Nomura}
\email{nomura@kias.re.kr}
\affiliation{School of Physics, KIAS, Seoul 02455, Korea}

\author{Hiroshi Okada}
\email{macokada3hiroshi@cts.nthu.edu.tw}
\affiliation{Physics Division, National Center for Theoretical Sciences, Hsinchu, Taiwan 300}

\date{\today}

\begin{abstract}
We  study a neutrino model introducing an additional nontrivial gauged lepton symmetry where 
the neutrino masses are induced at two-loop level while the first and second charged-leptons of the standard model are done at one-loop level.
As a result of model structure, we can predict one massless active neutrino, and there is a dark matter candidate.
Then we discuss neutrino mass matrix, muon anomalous magnetic moment, lepton flavor violations, oblique parameters, and  relic density of dark matter taking into account the experimental constraints.
\end{abstract}
\maketitle
\newpage

\section{Introduction}
A flavor dependent gauge model often plays an important role in explaining flavor specific measurements such as deviation from the standard model (SM) in semileptonic $B$ decay process $B\to K^{(*)}\mu^+\mu^-$ which is reported by LHCb~\cite{Aaij:2014ora,Aaij:2017vbb}, lepton flavor violations (LFVs) such as $\mu\to e \gamma$ process~\cite{TheMEG:2016wtm}, and muon anomalous magnetic moment($\Delta a_\mu$) at Brookhaven National Laboratory~\cite{Bennett:2006fi}. 
 Along thought of these ideas, the ref.~\cite{Crivellin:2016ejn} has established a flavor dependent model with a larger gauge group,
 and a smaller group such as $U(1)$ flavor dependent model can be found as a result of partially breaking the symmetry.
 
{In particular, it is interesting to consider active neutrino mass matrix with a lepton specific flavor dependent gauge symmetry $U(1)_L$ in the ref.~\cite{Crivellin:2016ejn}.
The active neutrino masses are not allowed at the tree level and it should be related to the $U(1)_L$ gauge symmetry breaking.
Then a radiative seesaw scenario is an attractive candidate to generate the mass, which can be achieved introducing some exotic particles. 
Furthermore we expect predictability in the active neutrino mass due to restriction from the gauge symmetry. 
}
 
 In this paper, we construct a neutrino model based on the lepton flavor symmetry $U(1)_L$, where the neutrino mass can be induced at two-loop level~\cite{2-lp-zB, Babu:2002uu, AristizabalSierra:2006gb, Nebot:2007bc, Schmidt:2014zoa, Herrero-Garcia:2014hfa, Long:2014fja, VanVien:2014apa, Aoki:2010ib, Lindner:2011it, Baek:2012ub, Aoki:2013gzs, Kajiyama:2013zla, Kajiyama:2013rla, Baek:2013fsa, Okada:2014vla, Okada:2014qsa, Okada:2015nga, Geng:2015sza, Kashiwase:2015pra, Aoki:2014cja, Baek:2014awa, Okada:2015nca, Sierra:2014rxa, Nomura:2016rjf, Nomura:2016run, Bonilla:2016diq, Kohda:2012sr, Dasgupta:2013cwa, Nomura:2016ask, Nomura:2016pgg, Liu:2016mpf, Nomura:2016dnf, Simoes:2017kqb, Baek:2017qos, Ho:2017fte, Nomura:2017xko, Guo:2017gxp, delAguila:2011gr}.
% Also we realize to fix the $U(1)_{\mu-\tau}$ symmetry starting from the general anomaly conditions.
{We first discuss the case of general charge assignment and the conditions to cancel gauge anomalies. 
Then phenomenological analysis is carried out by fixing the charge assignment for our particle contents. 
 As a result we predict one massless active neutrino, and discuss a dark matter candidate calculating relic density.
 In addition, we discuss lepton flavor violations (LFVs) and muon $g-2$ in the model.}

This paper is organized as follows.
In Sec.~II, we show the mechanism of anomaly cancellations, review our model setup, formulate the Higgs sector, fermion sector including active neutrinos,
muon anomalous magnetic moment, lepton flavor violations, and phenomenologies of a dark matter candidate where we show the allowed region of DM mass to satisfy the relic density without conflict of direct detection bound.
In Sec. III, we have globally numerical analysis, and investigate the allowed region to satisfy all the data that we will discuss.
Finally we devote Sec. IV  to conclusion.
%\newpage

%%%%%%%%%%%%%%%%%%%%%%%%%%%%%%%%%%%%%
%\section{The Model}
%\subsection{Model setup}

 \begin{widetext}
\begin{center} 
\begin{table}[t]%[tbc]
%\begin{tiny}
\begin{tabular}{|c||c|c|c|c|c|c|c||c|c|c|c|c|c|c|}\hline\hline  
%&\multicolumn{5}{c||}{SM leptons} & \multicolumn{3}{c|}{Exotic fermions} \\\hline
 & ~$L_{L_a}$~ & ~$L_{L_3}$~ & ~$e_{R_a}$~ &~$e_{R_3}$~ & ~$E_{L_a}$~ & ~$E_{R_a}$~ &~$E_{3}$~ & ~$H$~ & ~$\varphi_{a}$~ 
 & ~$\varphi_1$~ & ~$\eta$~  & ~$h^+$~ & ~$k^{++}$~& ~$s_0$~
\\\hline 
%$SU(3)_C$ & $\bm{3}$  & $\bm{3}$  & $\bm{3}$  & $\bm{1}$  & $\bm{1}$  & $\bm{1}$  & $\bm{1}$  & $\bm{1}$  \\\hline 
 %%%
 $SU(2)_L$ & $\bm{2}$  & $\bm{2}$  & $\bm{1}$ & $\bm{1}$ & $\bm{1}$  & $\bm{1}$& $\bm{1}$ & $\bm{2}$  & $\bm{1}$ & $\bm{1}$ 
 & $\bm{2}$  & $\bm{1}$ & $\bm{1}$& $\bm{1}$   \\\hline 
 %%%
$U(1)_Y$ & $-\frac12$ & $-\frac12$  & $-1$ & $-1$  & $-1$ & $-1$ & $-1$  & $\frac12$ & $0$ & $0$ & $\frac12$& $1$ & $2$ & $0$\\\hline
 %%%
 $U(1)_{L}$ & $x_a$ & $-1$  & $\frac12$ & $-1$  & $\frac12$ & $x_a$  & $-x_1-x_2$  & $0$   & $\frac12-x_a$   & $1$   & $0$ & $-\frac12$ & $-1$ & $0$ \\\hline
 $Z_2$ & $+$ & $+$ & $+$ & $+$ & $-$ & $-$ & $-$ & $+$ & $+$ & $+$ & $-$ & $-$ & $+$ & $-$ \\ \hline
 %%%
\end{tabular}
\caption{Field contents of fermions and bosons
and their generic charge assignments under $SU(2)_L\times U(1)_Y\times U(1)_{L}\times Z_2$, where $E_3$ is vector-like singly charged fermion, $a(=1-2)$ is flavor indices, and $x_a$ is nonzero arbitral charge.}
\label{tab:1}
% \end{tiny}
\end{table}
\end{center}
\end{widetext}

\section{Model setup and phenomenologies}
Here we construct our model with $U(1)_L$ symmetry and carry out phenomenological analysis.
In fermion sector, we introduce three exotic singly-charged fermions with different $U(1)_L$ charges;
$E_{L_a}$ and $E_{R_a}\ (a=1,2)$ respectively have $\frac12$ and $x_a$, while $E_3$, which is a vector-like fermion,   
has $-x_1-x_2$. 
Notice here that the charge of $E_3$ is arbitrary in general, but
$x_a$ are determined by several anomaly cancellations as discussed below.
%%%
In boson sector, we introduce three isospin singlet bosons with nonzero VEVs,
and 
%with different $U(1)_L$ charges;
$ \varphi_1$ and $\varphi_a\ (a=1,2)$ respectively have $1$ and $\frac12-x_a$.
In addition, we introduce a singly charged boson $h^\pm$, a doubly charged boson $k^{\pm\pm}$, and an isospin singlet(doublet) inert boson $s(\eta)$, and each of them has $U(1)_L$ charge of $-\frac12$, $-1$, and $0$. $H$ is identified as the SM-like Higgs.
%%%
Furthermore, we impose $Z_2$ odd for new fields except $\varphi_{1,a}$ and $k^{\pm\pm}$ in order to forbid the mixing between the SM fermions and exotic one and assure the stability of DM candidate; neutral component of $\eta$.
Thus neutrino masses are induced at two-loop level as shown below.
These particle contents and its charge assignments are summarized in Table~\ref{tab:1}.

%%%%%%%%% Anomaly cancellation %%%%%%%%%
{\it Anomaly cancellation}:
We explore conditions of anomaly cancellations under the  $SU(2)_L\times U(1)_Y\times U(1)_{L}$ gauge symmetry in table \ref{tab:1},
where $[U(1)_L]^2U(1)_Y$ is automatically zero under these symmetries. 
Thus one should consider three types of triangle anomalies including all the families as follows:
\begin{align}
& [SU(2)]^2 U(1)_L : \quad x_1 + x_2 -1 =0, \\
& [U(1)_Y]^2U(1)_L:\quad \frac12(x_1+x_2)-\frac12=0,\\
%& [U(1)_L]^2U(1)_Y:\quad 0,\\
& [U(1)_L]^3:\quad -(x_1^3+x_2^3)+1=0,\\
& U(1)_L: \quad -(x_1+x_2)+1=0.
\end{align}
Therefore one finds the following two conditions:
\begin{align}
x_1+x_2=1,\quad x_1^2-x_1x_2+x_2^2=1.
\end{align}
One finds several simple solutions such as $(x_1,x_2)=(0,1),(1,0)$ that respectively correspond to $\mu-\tau$ and $e-\tau$ symmetry for {lepton doublets.}
%%%%%%%%% %%%%%%%%% %%%%%%%%%

{\it Fixing charge assignments}: Here we fix to be $x_1=0$ and $x_2=1$ for simplicity. %that reach an $U(1)_{\mu-\tau}$ gauge symmetry. 
In this case, one can simply define $\varphi_a\equiv \varphi_0$ with  $U(1)_L$ charge of $\frac12$.
Under these symmetries in table~\ref{tab:1}, the renormalizable Lagrangian in the lepton sector and Higgs potential are respectively given by 
{\begin{align}
%\label{eq:tri-y}
&-{\cal L}_{\rm lepton}  = y_{\ell_3} \bar L_{L_3} H e_{R_3} + f_{\alpha} \bar E_{R_\alpha}\tilde \eta L_{L_\alpha}
+g_{L_{ab}} \bar E^c_{L_a} E_{L_b} k^{++}  + \kappa_a \bar E_{L_a} e_{R_a} s_0 \label{eq:nontri-y-1} \\
&
+y_{\varphi_{a1}} \bar E_{L_a} E_{R_1} \varphi_0 +y_{\varphi_{a2}}\bar E_{L_a} E_{R_2} \varphi_0^* 
+y_{\varphi_{31}}\bar E_{L_3} E_{R_1} \varphi_1^* 
+ M_{33} \bar E_{L_3} E_{R_3}   \label{eq:nontri-y-2}  \\
&
+ g_{ab} \bar e^c_{R_a} e_{R_b} k^{++} + g_{R_{12}} (\bar E^c_{R_1} E_{R_2}+ \bar E^c_{R_2} E_{R_1}) k^{++}+{\rm c.c.} 
  \label{eq:nontri-y-3} , 
  \end{align}
%%% %%%
\begin{equation}
V_{\rm non-trivial}= \mu [H^\dag \eta s_0 +{\rm c.c.}] 
+\frac{\mu_{khh}}2 [k^{++} (h^-)^2 +{\rm c.c.}]  + \frac{\mu_{\varphi}}2 [\varphi_0^2\varphi_1^* +{\rm c.c.}] 
 + \lambda_0 [(H^Ti\sigma_2 \eta)h^- \varphi_0^*+{\rm c.c.}] 
\label{eq:lag-lep}
\end{equation}
}
where  $\alpha$ runs over $1$ to $3$, and $(a,b)$ run over $1$ to $2$,
$\tilde H \equiv (i \sigma_2) H^*$ with $\sigma_2$ being the second Pauli matrix.
The first term gives the masses for the SM charged-leptons, and $f,g_{L/R}$ as well as $ \mu_{khh}$ and $\lambda_0$
contribute to the structure of neutrino masses. $\mu_\varphi$ forbids the massless goldstone boson (GB) arising from $\varphi_{0,1}$.
On the other hand $g_{ab}$ does not contribute to the neutrino masses and $g_{R_{12}}$ itself cannot reproduce the experimental results for neutrinos.
Thus we just neglect these terms in our analysis.

%%%%%%%%%
%\subsection{Scalar sector and $Z'$ boson}
%{\it Scalar sector}:　
Next we formulate the scalar sector, in which we parameterize as follows: 
\begin{align}
%\begin{tiny}
&H =\left[\begin{array}{c}
w^+\\
\frac{v + h +i z}{\sqrt2}
\end{array}\right],\quad 
%%%
\eta =\left[\begin{array}{c}
\eta^+\\
 \frac{\eta_0 + i \eta_I}{\sqrt2} %+ \frac{i \eta_I}{\sqrt2}
\end{array}\right],\quad 
%%%
\varphi_{0,1}=
\frac{v'_{0,1}+\sigma_{0,1} + iz'_{\varphi_{0,1}}}{\sqrt2},
\label{component}
%\end{tiny}
\end{align}
where $w^+$, and two massless eigenstates among $z$, and $z'_{\varphi_{0,1}}$ are absorbed by the SM and $U(1)_L$  gauge bosons $W^+$, $Z$, and $Z'$.
%%%
Inserting tadpole conditions, the CP even mass matrix in basis of $(h,\sigma_0, \sigma_1)$
 can be formulated by $O_R M_R^2 O_R^T\equiv {\rm Diag}[m_{h_1}^2,m_{h_2}^2, m_{h_3}^2]$, where $h_1$ is the SM Higgs and $O_R$ is three by three orthogonal mixing matrix.
%%%
{On the other hand the inert boson mass matrix in basis of $(\eta_0,s_0)$ is formulated by $U M_{inert}^2 U^T\equiv {\rm Diag}[m_{\eta}^2,m_s^2]$, where $U$ is two by two orthogonal matrix.
Then one can parametrize the relation between flavor and mass eigenstate as~\cite{Okada:2014qsa} 
\begin{align}
\left[\begin{array}{c}\eta_0 \\  s_0 \\ \end{array}\right]=  %
\left[\begin{array}{cc}\cos\beta &- \sin\beta \\
  \sin\beta & \cos\beta \\ \end{array}\right]
\left[\begin{array}{c}\eta \\  s \\ \end{array}\right]
 ,\end{align}
 where $\sin\beta$ is proportional to $\mu v$, and we expect to be $\sin\beta<<1$ in our analysis below.
 }
%%%
On the other hand the singly charged boson mass matrix in basis of $(\eta^\pm,h^\pm)$ is formulated by $O_\pm M_\pm^2 O_\pm^T\equiv {\rm Diag}[m_{H_1^\pm}^2,m_{H_2^\pm}^2]$, where $O_\pm$ is two by two orthogonal matrix.~\footnote{In general it should be unitary. But it can be orthogonal when all the coupling of Higgs potential are real.}
Then one can parametrize the relation between flavor and mass eigenstate as~\cite{Okada:2014qsa} 
\begin{align}
\left[\begin{array}{c}\eta^\pm \\  h^\pm \\ \end{array}\right]=  %
\left[\begin{array}{cc}\cos\theta &- \sin\theta \\
  \sin\theta & \cos\theta \\ \end{array}\right]
\left[\begin{array}{c}H_1^\pm \\  H_2^\pm \\ \end{array}\right]
 ,\end{align}
 where $\sin\theta$ is proportional to $\lambda_0 vv'_0$.
%where $D\equiv \sqrt{(v'^2 \lambda_{\varphi} - v^2 \lambda_{H})^2 + (vv'\lambda_{H\varphi})^2} $, and $s_{\alpha}(c_{\alpha})$ is the short hand notation of $\sin\alpha(\cos\alpha)$.
  
  %%%%%%%%%%
  {
 % {\it $Z'$ boson}: ...
  \if0
  After spontaneous $U(1)_L$ gauge symmetry breaking, we have $Z'$ boson that couples not to quarks but leptons.
  The mass of $Z'$ is given by $m_{Z'} = 4 g' v'$ where $g'$ is the $U(1)_L$ gauge coupling.
  Since our $Z'$ universally couples to SM leptons the LEP experiment provides the strongest constraints on the gauge coupling and $Z'$ mass.
  Assuming $m_{Z'} \gtrsim 200$ GeV, the LEP constraint is applied to the effective Lagrangian 
\begin{equation}
L_{eff} = \frac{1}{1+\delta_{e \ell}} \frac{g'^2}{m_{Z'}^2} (\bar e \gamma^\mu e)( \bar \ell \gamma_\mu \ell)
\end{equation}
where $\ell = e$, $\mu$ and  $\tau$.
We then obtain following constraint from the analysis of data by measurement at LEP~\cite{Schael:2013ita}: 
%which  tells us the following restriction:
\begin{align}
\frac{m_{Z'}}{g'} \gtrsim 7.0\ {\rm TeV}.\label{eq:lep}
\end{align}
We will take into account this constraint in the analysis of DM relic density and collider physics below.
\fi
 } 
  
{
{\it  $Z'$ boson}: A massive $Z'$ boson appears after spontaneous symmetry breaking of $U(1)_L$. The mass of $Z'$ is given by $m_{Z'} = g' \sqrt{v_0'^2 + 4 v_1'^2}/2$ where $g'$ is gauge coupling of $U(1)_L$. Since the $Z'$ couples to electrons the mass and gauge coupling are constrained by the LEP data. Here we simply assume the mass is around TeV scale and the value of $g'$ satisfies the constraints.  Note that $Z'$ does not contribute to neutrino mass generation and our DM candidate has no direct interaction with the $Z'$ since $\eta$ does not have $U(1)_L$ charge. Thus we will not discuss $Z'$ physics in our analysis.
  }
  
  %%%%%%%%%%%
{\it Exotic charged-fermion masses}:
After the symmetry breaking, the exotic charged fermion mass matrix in Eq.(\ref{eq:nontri-y-2}) can be given in the basis $[E_{1},E_{2},E_{3}]^T$ as follows:
\begin{align}
M_{E}\equiv \left[\begin{array}{ccc}  M_{11} & M_{12} & M_{13}  \\ M_{12} & M_{22} & 0 \\ M_{13} & 0 & M_{33} \end{array}\right],\label{eq:ML'}
\end{align}
where we {have assumed $M_{E}$ to be a real symmetric matrix} for simplicity and define $M_{11}\equiv y_{\varphi_{11}} v'_0/\sqrt2$, $M_{12}\equiv y_{\varphi_{21}} v'_0/\sqrt2$, $M_{22}\equiv y_{\varphi_{22}} v'_0/\sqrt2$, and $M_{13}\equiv y_{\varphi_{31}} v'_1/\sqrt2$. Then $M_{E}$ is diagonalized by an orthogonal mixing matrix $V$ ($VV^T=1$) as 
\begin{align}
V M_E V^T =D_E \equiv {\text{Diag.} }\left[M_e,M_\mu,M_\tau\right],\quad E_{1,2,3}=V^T E_{e,\mu,\tau},\label{eq:N-mix}
\end{align}
where $M_{e,\mu,\tau}$ is the mass eigenstate.
%%%

%%%%%%%%%%%%%%%%%%%
\begin{figure}[t]
\begin{center}
\includegraphics[width=80mm]{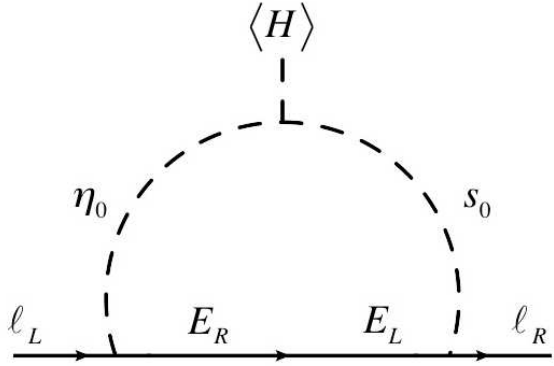} 
\caption{The one loop diagram to generate SM charged lepton masses. } 
  \label{fig:chgd-lp}
\end{center}\end{figure}
%%%%%%%%%%%%%%%%%%%

{
{\it SM charged-fermion masses}:
Since the first and second charged-leptons are not induced at the tree level, but done at the one-loop level in  fig.~\ref{fig:chgd-lp}.
In order to formulate these masses, let us write down the relevant Lagrangian to the SM charged-leptons in the mass eigenbasis as
\begin{align}
- {\cal L}&\sim 
\sum_i^{1-3} \sum_{\alpha}^{e,\mu,\tau} {{V_{\alpha i} f_i }}
\bar E_{R_\alpha}  \ell_{L_i} (\cos\beta \eta - \sin\beta s) 
+
\sum_{j}^{1,2}  \sum_{\alpha}^{e,\mu,\tau} { V_{\alpha j} \kappa_j }
\bar E_{L_\alpha}  \ell_{R_j} (\sin\beta \eta + \cos\beta s) 
+
m_{\ell_{33}} \bar e_{L_3} e_{R_3} %+{\rm c.c.}
\nn\\
%%%
&\equiv
\sum_i^{1-3} \sum_{\alpha}^{e,\mu,\tau} {F_{\alpha i}}
\bar E_{R_\alpha}  \ell_{L_i} (\cos\beta \eta - \sin\beta s) 
+
\sum_{j}^{1,2}  \sum_{\alpha}^{e,\mu,\tau} { H_{\alpha j} }
\bar E_{L_\alpha}  \ell_{R_j} (\sin\beta \eta + \cos\beta s) 
+
m_{\ell_{33}} \bar e_{L_3} e_{R_3},
\end{align}
where $m_{\ell_{33}}\equiv y_{\ell_3} v/\sqrt{2}$.
%is the rank {\color{blue}one} matrix. 
Then the mass matrix for the charged-leptons can be induced as follows~\cite{Nomura:2016emz, Nomura:2016pgg, Nomura:2017ezy}:
\begin{align}
& (m_{\ell})_{ab} =\sum_{j}^{1,2} \sum_i^{1-3}  (\delta m_\ell)_{ji}+ m_{\ell_{33}} ,\\
%%%
& (\delta m_\ell)_{ji} =\frac{\sqrt2 \sin \beta \cos \beta}{(4\pi)^2} \sum_{\alpha}^{e,\mu,\tau}(H^\dag)_{j\alpha} M_\alpha F_{\alpha i}
\int_0^1dx \frac{x M^2_\alpha+(1-x) m_{s}^2}{x M^2_\alpha+(1-x) m_{\eta}^2}.
\end{align}
 $(m_{\ell})_{ab}$ is generally diagonalized by bi-unitary matrices as {$V_{\ell_R} (m_{\ell})_{ab} V_{\ell_L}^\dag\equiv D_\ell$, where $D_\ell$ is mass eigenstate of charged-leptons.
 Then the resulting mass eigenvalues for the SM charged-leptons are generally given by
\begin{align}{V_{\ell_L}  m_{\ell}^\dag m_{\ell}  V_{\ell_L}^\dag} ={\rm diag.}\left(|m_e|^2,|m_\mu|^2,|m_\tau|^2\right),\end{align}
}
Thus the observed lepton mixing arises from the neutrino part only.  
}
%%%

%%%%%%%%%%%%%%%%%%%
\begin{figure}[t]
\begin{center}
\includegraphics[width=80mm]{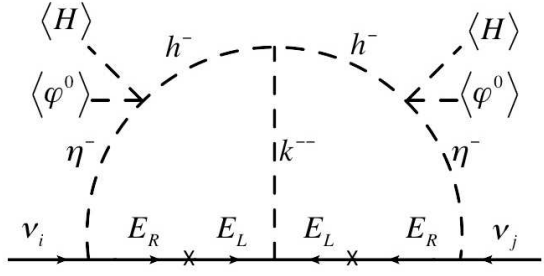} 
\caption{The two loop diagram to generate neutrino masses. } 
  \label{fig:neut}
\end{center}\end{figure}
%%%%%%%%%%%%%%%%%%%

{\it Active neutrinos}:
First of all, let us write down the relevant Lagrangian to the neutrinos in the mass eigenbasis as
\begin{align}
- {\cal L}&\sim 
\sum_{\alpha=e,\mu,\tau} { V_{\alpha i}f_i }
\bar E_\alpha P_L \nu_i (\cos\theta H_1^- - \sin\theta H_2^-)
+\sum_{a,b=1}^2 V_{ia} (g_L)_{ab}(V^T)_{bj} \bar E^c_{L_i} E_{L_j} k^{++} +{\rm c.c.}\nn\\
%%%
&\equiv
 \sum_{\alpha=e,\mu,\tau} { F_{\alpha i}} \bar E_\alpha P_L\nu_i (\cos\theta H_1^- - \sin\theta H_2^-)
+(G_L)_{ij} \bar E^c_{L_i} E_{L_j} k^{++} +{\rm c.c.},
\end{align}
where $g_L$ is the rank two matrix. 
%%%
Then the neutrino mass matrix is induced at the two-loop level in fig.~\ref{fig:neut}, which
 is given by~\cite{Okada:2014qsa}
\begin{align}
& (m_\nu)_{ij}\approx 
\frac{\mu_{khh} \sin^22\theta}{2(4\pi)^4}\sum_{\alpha,\beta=e,\mu,\tau}
\left[\frac{{ F_{i\alpha}^T} M_\alpha (G_L)_{\alpha\beta} M_{\beta} {F_{\beta j}} }{M^2_\alpha} \right] \times \\
%%%
&
\left[ {\cal F_I} (r_{H_1}^\alpha,r_{k}^\alpha,r_{H_1}^\alpha,r_{\alpha}^\beta) - {\cal F_I} (r_{H_2}^\alpha,r_{k}^\alpha,r_{H_1}^\alpha,r_{\alpha}^\beta) 
-{\cal F_I} (r_{H_1}^\alpha,r_{k}^\alpha,r_{H_2}^\alpha,r_{\alpha}^\beta)+{\cal F_I} (r_{H_2}^\alpha,r_{k}^\alpha,r_{H_2}^\alpha,r_{\alpha}^\beta)  \right],\nn\\
& {\cal F_I} (r_1,r_2,r_3,r_4)\equiv \int_0^1\frac{[dx][dx']}{z(z-1) (x'+y' r_1)-z'(x r_4+yr_3+zr_2)}, 
\end{align}
where $r_{\beta}^\alpha\equiv \frac{M_\beta^2}{M_\alpha^2}$, $[dx]\equiv dxdydz\delta(x+y+z-1)$, and $[dx']\equiv dx'dy'dz'\delta(x'+y'+z'-1)$.
%%%
Whereas we also formulate the experimental neutrino mass matrix as $m_\nu^{\exp.} \approx |V_{MNS}^\dag D_\nu  V_{MNS}^*|$
that
can be determined by neutrino oscillation data, when numerical (Dirac and Majorana) phases are provided.
{\it Notice here that one of the three active neutrino mass eigenstates is massless, since the matrix rank $g_L$ is two; $|m_{\nu_{1(3)}}|=0$ in case of normal(inverted) hierarchy. The structure of the mass matrix indicate that the neutrino mixing angles mainly arise from the mixing matrix $V$.}
%%% 
Then one finds the following ranges at 3$\sigma$ confidential level~\cite{Forero:2014bxa} given by~\footnote{Recently $\delta=-\pi/2$ is experimentally favored.
 But our result does not change significantly, 
even if we fix to be $\delta=-\pi/2$.}, assuming the normal one:
\begin{align}
%({\rm NH}):\
{m}_\nu^{exp.}&\approx \left[\begin{array}{ccc} 
0.0845-0.475 & 0.0629-0.971 &0.0411-0.964 \\
* & 1.44-3.49 &  1.94-2.85 \\
* & * &   1.22-3.33\\
  \end{array}\right]\times 10^{-11}\ {\rm GeV},\label{eq:exp-Neutmass-NH}
%\\
%V_{MNS}&=\left[\begin{array}{ccc} {c_{13}}c_{12} &c_{13}s_{12} & s_{13} e^{-i\delta}\\
% -c_{23}s_{12}-s_{23}s_{13}c_{12}e^{i\delta} & c_{23}c_{12}-s_{23}s_{13}s_{12}e^{i\delta} & s_{23}c_{13}\\
%  s_{23}s_{12}-c_{23}s_{13}c_{12}e^{i\delta} & -s_{23}c_{12}-c_{23}s_{13}s_{12}e^{i\delta} & c_{23}c_{13}\\
%  \end{array}\right]
  %%%
  %\left[\begin{array}{ccc} e^{i\alpha_1/2} & 0 &0 \\0 & e^{i\alpha_2/2} & 0 \\0 & 0 &  1\\  \end{array}\right],
\end{align}
and Majorana phases $\alpha_{1,2}$ taken to be $\alpha_{1,2}\in[-\pi,\pi]$.
In the numerical analysis, we impose the constraint $|{m}_\nu^{}|\approx {m}_\nu^{exp.}$.

%%%%%%%%%
 {\it Muon anomalous magnetic moment ($\Delta a_\mu$)}:
%{\color{red} (please revise this part)}:
%%%
 $\Delta a_\mu$ has been observed and its discrepancy is estimated by~\cite{Hagiwara:2011af}
\begin{align}
\Delta a_\mu=(26.1\pm8.0)\times10^{-10}.\label{eq:exp-g2mu}
\end{align}
%%%
The relevant Yukawa Lagrangian contributing to $\Delta a_\mu$ as well as LFVs in the mass eigenbasis is given by
{ \begin{align}
- {\cal L}&\sim 
\sum_{i}^{1-3}  \sum_{\alpha=e,\mu,\tau}  F_{\alpha i} \bar E_\alpha P_L \ell_i \eta
+
\sum_{j}^{1,2}  \sum_{\alpha}^{e,\mu,\tau} {H_{\alpha j} } \bar E_{L_\alpha}  \ell_{R_j}  s
+ {\rm c.c.},
\label{eq:lfv-g2}
\end{align}
where $\eta\equiv \eta_R\simeq \eta_I$ and $\sin\beta<<1$.
Then our $\Delta a_\mu$ is induced by interaction with $F_{}$ coupling as explained above, and its form is computed as
\begin{align}
\Delta a_{\mu}^Y&\approx \frac{2m_\mu^2}{(4\pi)^2} \sum_{\alpha=e,\mu,\tau}
\left(|F_{2\alpha}|^2 F_{II}(m_{\eta}, M_{\alpha}) + |H_{2\alpha}|^2 F_{II}(m_{s}, M_{\alpha}) \right),
\\
%%%
F_{II}(m_a,m_b)&\equiv\frac{2 m_a^6+3m_a^4m_b^2-6m_a^2m_b^4+m_b^6+12m_a^4m_b^2\ln\left[\frac{m_b}{m_a}\right]}{12(m_a^2-m_b^2)^4}.
%,
\end{align}
%where one finds $F_{II}={\cal O}(0.1)$.
Considering the neutrino oscillations and lepton flavor violations for $F$ term as will be discussed below, we find the maximal order of $\Delta a_\mu^Y$ to be $10^{-12}$. On the other hand the term with $H_{2\alpha}$ provides the dominant contribution to $\Delta a_\mu$, since it is not constrained by any phenomenologies once we take $H_{21},H_{23}<<H_{22}$.
\\
%%%
In addition, $Z'$ gauge boson can contribute to $\Delta a_\mu$ and its form is approximately given by
\begin{align}
 \Delta a_{\mu}^{Z'}\approx 
 \frac{g_{Z'}^2 m_\mu^2}{48\pi^2 M_{Z'}^2},
 %\frac{g_{Z'}^2}{8\pi^2}\int_0^1 da \frac{2 r a (1-a)^2}{r(1-a)^2+a}, 
 \label{eq:G2-ZP}
\end{align}
where %$r\equiv(m_\mu/M_{Z'})^2$, and 
$Z'$ is the new gauge vector boson. %and we take the limit of $m_\mu/M_{Z'}<<1$.
Since the right-handed electron couples to the $Z'$ boson, we have the constraint $3.7\ {\rm TeV}\lesssim M_{Z'}/g_{Z'}$~\cite{Nomura:2017tih};  the maximal value of $\Delta a_\mu^{Z'}$ is $\Delta a_\mu^{Z'}({\rm Max})=1.72\times10^{-12}$.
%%%
Combining $\Delta a_{\mu}^Y$ and $\Delta a_\mu^{Z'}$,
we find the final result of muon $g-2$; $\Delta a_\mu\equiv \Delta a_{\mu}^Y+\Delta a_\mu^{Z'}$,
where we will adapt the maximal value $\Delta a_\mu^{Z'}({\rm Max})$ in our numerical analysis below.

%one finds our result in terms of $f$ in fig.~\ref{fig:ms-damu},

{\it Lepton flavor violations (LFVs)}: LFV processes of $\ell \to \ell' \gamma$ are given by the same term as the $(g-2)_\mu$, and their forms are given by
\begin{align}
BR(\ell_i\to \ell_j \gamma)
&\approx\frac{48\pi^3 C_{ij} \alpha_{em} }{(4\pi)^4G_F^2}
\left|\sum_{\alpha=e,\mu,\tau} F^\dag_{j\alpha} F_{\alpha i} F_{II}(m_{\eta^0}, M_{\alpha})\right|^2,
\label{eq:lfvs}
\end{align}
where $\alpha_{em}\approx1/137$ is the fine-structure constant, $G_F\approx1.17\times10^{-5}$ GeV$^{-2}$ is the Fermi constant,
and $C_{21}\approx1$, $C_{31}\approx 0.1784$, $C_{32}\approx0.1736$. 
Experimental upper bounds are given by~\cite{TheMEG:2016wtm, Adam:2013mnn}: 
\begin{align}{\rm BR}(\mu\to e \gamma)\lesssim 4.2\times 10^{-13},\ 
{\rm BR}(\tau\to e \gamma)\lesssim 3.3\times 10^{-8},\ 
{\rm BR}(\tau\to \mu \gamma)\lesssim 4.4\times 10^{-8},\label{eq:lfvs-cond}
\end{align}
where we define $\ell_1\equiv e$,  $\ell_2\equiv \mu$, and  $\ell_3\equiv \tau$. 
%This also gives a simple bound on the Yukawa coupling as follow:
%\begin{align}\left| \sum_{\alpha=e,\mu,\tau} F^\dag_{j\alpha} F_{\alpha i} F_{II}(m_{\eta^0}, M_{\alpha})\right|
%\lesssim%(4\pi)^2 \cdot10\cdot {\rm G_F} \times\sqrt{\frac{{\rm BR}(\ell_i\to \ell_j \gamma)}{48\pi^3C_{ij} \alpha_{\rm em} }}.\label{eq:lfvs-cond}\end{align}

{\it Oblique parameter}:
Since we have an isospin doublet boson $\eta$, we have to consider the oblique parameter known as $\Delta S$ and $\Delta T$~\cite{Barbieri:2006dq}.  In our case, one finds the following relations:
\begin{align}
\Delta S\approx \frac{1}{12\pi}\ln\left[\frac{m^2_{\eta}}{m_{H_1^\pm}^2}\right],\
\Delta T\approx \frac{1}{16\pi^2 v^2 \alpha_{\rm em}}
\left[\frac{m^2_{\eta}+m_{H_1^\pm}^2}{2}
-
\frac{m_{H_1^\pm}^2 m^2_{\eta}}{m_{H_1^\pm}^2- m^2_{\eta}} \ln\left[\frac{m_{H_1^\pm}^2}{m^2_{\eta}}\right] \right],
%\label{eq:st-cond}
\end{align}
where let us remind the conditions $(\theta,\beta)<<1$; $m_{\eta^\pm}\approx m_{H^\pm_1}$ and $m_\eta \equiv m_{\eta_R} \approx m_{\eta_I}$.
Then the current bounds are given by~\cite{Cheung:2017efc}
\begin{align}
-0.04\lesssim \Delta S\lesssim 0.14,\quad
0.01\lesssim \Delta T\lesssim 0.15.
\label{eq:st-cond}
\end{align}
%%%

%%%%%%%%%%%%%%%%%%%
\begin{figure}[t]
\begin{center}
\includegraphics[width=100mm]{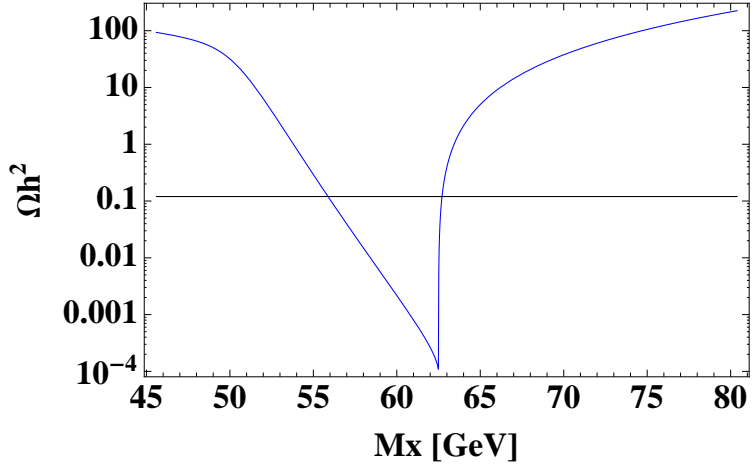} 
\caption{Plot line of relic density in term of the DM mass, where we have used $\lambda_{XXh_1}^{\rm Max}=0.01$. 
Horizontal line represents the measured relic density $\sim$0.12.} 
  \label{fig:relic-dm}
\end{center}\end{figure}
%%%%%%%%%%%%%%%%%%%
{\it Dark matter}:
{In our model, we have two types of  DM candidates; $\eta$ and $s$.
But let us here focus on the neutral component of $\eta$ can be DM candidate resymbolized by $X\equiv \eta$, because $\eta$ is more testable than $s$ due to constraining the mass than $s$ from experiments such as neutrino oscillations, LFVs, and oblique parameters.}
%%%
Although the general analysis has been done by Ref.~\cite{Hambye:2009pw}, we impose a constraint of  the DM mass, which is smaller than the mass of $W$ boson, but greater than the half of Z boson mass to forbid the invisible decay of Z boson;
therefore
 \begin{align}
\frac{m_Z}2\lesssim M_X\lesssim m_W.
 \end{align}
This region is in favor of getting sizable muon $g-2$, 
and well testable in  the direct detection constraint such as LUX experiment~\cite{Akerib:2016vxi} because it provides the most severe bound at around 50 GeV. 
%as well as evading  the direct detection constraint such as LUX experiment. 
%Also we note that annihilation modes from Higgs portal is subdominant when we require to which is discussed below.
%%%
Under the condition, we have two relevant annihilation cross sections to explain the relic density of DM.
One mode arises from Yukawa coupling $F$ that gives the d-wave dominance, and another one does from s-channel via SM Higgs with final state of bottom pairs, where we assume mixing among the CP-even neutral bosons are negligible. 
%%%
The $d-$wave dominant cross section given by Eq.~(\ref{eq:lfv-g2}) is found to be~\cite{Das:2017ski}
\begin{align}
\sigma v_{\rm rel}\approx 
\frac{ M_X^6}{60\pi}
\sum_{i,j = 1}^3\left|\sum_{\alpha=e,\mu,\tau} \frac{F^\dag_{i\alpha} F_{\alpha j}}{ (M_{\alpha}^2+M_X^2)^2} \right|^2.
%v_{\rm rel}^4 \equiv d_{\rm eff} v_{\rm rel}^4.
\end{align}
In our estimation, however,  this cross section reaches $10^{-10}$ [GeV]$^{-2}$ at most, which is smaller than the cross section required to give right relic density by one order of magnitude. Thus we have to rely on Higgs portal interaction mode, and its dimensionless cross section $W$ is found to be
\begin{align}
&W(s)=\frac{3  \lambda_{XXh_1}^2 m_b^2 (s-4 m_b^2)} {4\pi |s-m_{h_1}^2 + i m_h \Gamma^{h_1}_{\rm tot}|^2}\sqrt{1-\frac{4 m_b^2}{s}},\\
%%%
&\Gamma^{h_1}_{\rm tot}\equiv \Gamma^{h_1}_{\rm SM} + \Gamma^h_{\rm inv},\quad 
\Gamma^h_{\rm inv}= \frac{\mu_{XXh_1}^2 v^2}{64\pi} \sqrt{1-\frac{4 M_X^2}{m_{h_1}}}
\Theta\left(\frac{m_{h_1}}{2} - M_X\right),
\end{align}
where $\lambda_{XXh_1}(\equiv \lambda_{H\eta}+\lambda'_{H\eta}+2\lambda''_{H\eta})$ is the trilinear couplings of $X-X-h_1$, arising from $ \lambda_{H\eta} |\eta|^2|H|^2+ \lambda'_{H\eta}|\eta^\dag H|^2+ \lambda''_{H\eta}/2(\eta^\dag H)^2+{\rm h.c.}$. Notice here that $\mu_{XXh_1}$ is restricted by the direct detection with spin independent scattering via Higgs portal,~\footnote{The constraint of invisible decay of the SM Higgs always gives milder than the one of  direct detection in our parameter region. Thus we will not discuss here.} and its bound is conservatively found to be~\cite{Das:2017ski} 
\begin{align}
\lambda_{XXh_1}\lesssim 0.01.\label{eq:lamxxh}
\end{align}

Here we apply the following formula to get the relic density of DM given by~\cite{Edsjo:1997bg};
\begin{align}
&\Omega h^2
\approx 
\frac{1.07\times10^9}{\sqrt{g_*(x_f)}M_{Pl} J(x_f)[{\rm GeV}]},
\label{eq:relic-deff}
\end{align}
where $g^*(x_f\approx25)$ is the degrees of freedom for relativistic particles at temperature $T_f = M_X/x_f$, $M_{Pl}\approx 1.22\times 10^{19}$ GeV,
and $J(x_f) (\equiv \int_{x_f}^\infty dx \frac{\langle \sigma v_{\rm rel}\rangle}{x^2})$ is given by~\cite{Nishiwaki:2015iqa}
\begin{align}
J(x_f)&=\int_{x_f}^\infty dx\left[ \frac{\int_{4M_X^2}^\infty ds\sqrt{s-4 M_X^2} W(s) K_1\left(\frac{\sqrt{s}}{M_X} x\right)}{16  M_X^5 x [K_2(x)]^2}\right], \label{eq:relic-deff}
\end{align}
%%%
where $M_P\approx 1.22\times10^{19}$ GeV is the Planck mass, $g_*\approx 100$ is  the total number of effective relativistic degrees of freedom at the time of freeze-out, and $x_f\approx25$ is defined by $M_X/T_f$ at the freeze out temperature ($T_f$).
Then one has to satisfy the the current relic density of DM; $\Omega h^2\approx 0.12$~\cite{Ade:2013zuv}.
In fig.~\ref{fig:relic-dm}, we show the line of $\Omega h^2$ in terms of $M_X$, where we have used the maximum value in Eq.(\ref{eq:lamxxh}); $\lambda_{XXh_1}^{\rm Max}=0.01$. 
Thus one finds that the resulting allowed region is 
\begin{align}
54\ {\rm GeV}\lesssim M_X\lesssim 62.5\ {\rm GeV}, \ {\rm for}\  \lambda_{XXh_1}\lesssim 0.01,
\label{eq:dm-cond}
\end{align}
where the upper bound of the DM mass; $62.5\ {\rm GeV}$, arises from the pole mass of the half SM Higgs.
%It suggests that the allowed point is at around the half mass of the SM Higgs, and inside the line represents the region of $\lambda_{XXh_1}< 0.02$.
%%%

%%%%%%%%%%%%%%%%%%%
\begin{figure}[t]
\begin{center}
\includegraphics[width=100mm]{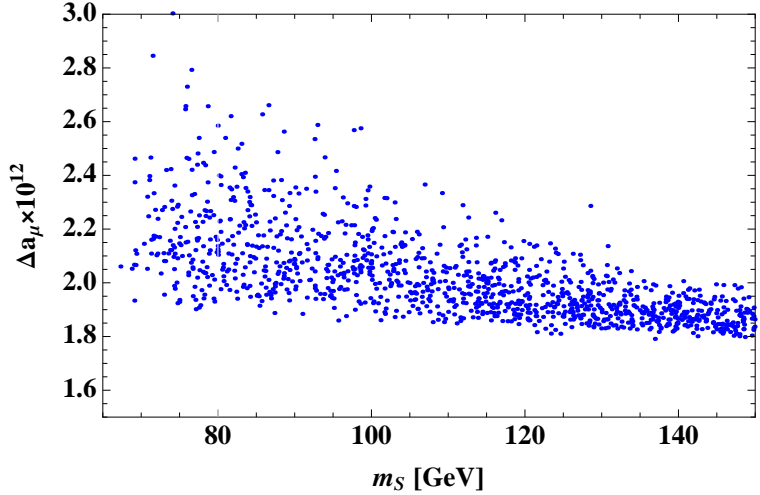} 
\caption{Scattering plots on muon $g-2$ and $m_S$ plane where each point satisfies experimental constraints.} 
  \label{fig:ms-damu}
\end{center}\end{figure}
%%%%%%%%%%%%%%%%%%%
\begin{figure}[t]
\begin{center}
\includegraphics[width=75mm]{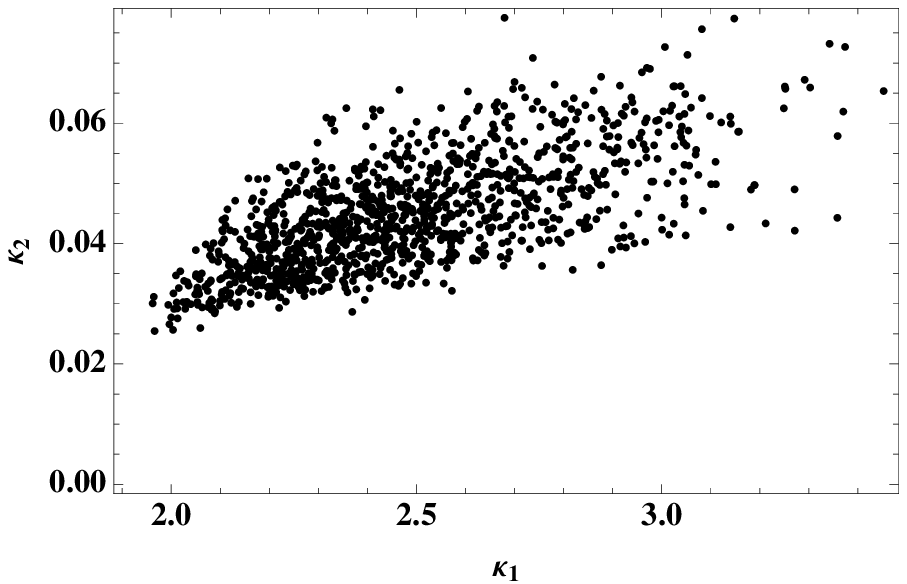} \quad
\includegraphics[width=75mm]{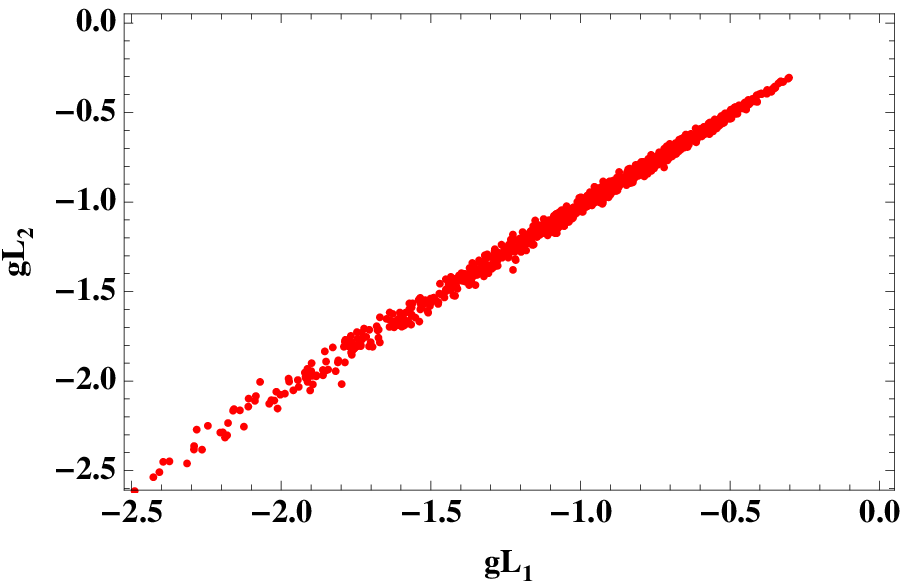} 
\caption{Left plot: allowed parameter points on $k_1$-$k_2$ plane. Right plot: allowed parameter points on $g_{L_1}$-$g_{L_2}$ plane.} 
\label{fig:yukawa}
\end{center}\end{figure}
%\section{Numerical analysis}
%where  red, green, and blue respectively present the case of $f_{22}$, $f_{11}$, and $f_{33}$. It suggests that the maximal value of muon $g-2$ is of the order $3\times10^{-12}$ and typical scale of $f$ is about $10^{-2}$, which is  three order of magnitude below than the experimental result.

\section{Numerical analysis}
In this section, we show a global analysis.
%where we have fixed some parameters for simplicity.
Before the numerical analysis, we work on the diagonal basis of $g_L$ by the phase redefinition of $E_L$; $g_{L}=$Diag.($g_{L_1},g_{L_2},g_{L_3}$).
Then we directly solve the couplings $\kappa_{1,2}$ and $g_{L_{1,2}}$ by using the relations $|V_{\ell_R}m_\ell V^\dag_{\ell_L}|_{11(22)}=|m_{e(\mu)}|$ and $|{m}_\nu|_{11(12)}|\approx (m_\nu^{exp.})_{11(22)}$, respectively, where we impose the perturbative bounds on these output parameters; $(\kappa_{1,2},\ g_{L_{1,2}})\lesssim \sqrt{4\pi}$.
\footnote{In principle, all the Yukawa couplings could be solved by using all the components of these relations. However it is technically difficult in our model.}
%%%
Now we randomly select the following range of reduced input parameters as
\begin{align}
& M_E\in[100,1000]\ {\rm GeV},  \ m_{H_1^\pm} \in [200,500]\ {\rm GeV},  \ 
{m_{H_2^\pm} \in [80,M_X+140]\ {\rm GeV}}, \ m_{k^{++}} \in [500,600]\ {\rm GeV},
\nn\\
%%%
& \mu_{khh}\in[1,2]\ {\rm GeV},  \ (\sin\theta,\sin\beta) \in [0,\pi/4], \ m_{k^{++}} \in [500,600]\ {\rm GeV},
 \ m_{S} \in [1.2 M_X,150]\ {\rm GeV},
\nn\\
%%%
& |f_{e,\mu,\tau},\kappa_3,g_{L_{3}}|  \in[0.001, 1],\  (\rho,\sigma)  \in[0, \pi],\  \delta \in[\pi, 2\pi],\  |s_\alpha| \in[10^{-5},0.1],
 %\\%%%&  M_{Z'}\in [10^{-3},10^3]\ [{\rm GeV}],\  g_{Z'} \in[10^{-5},10^{-3}],
\end{align}
{where the lower mass range for $m_{H^\pm_2}$ arises from the bound from {LEP data~\cite{Abbiendi:2008aa}}, while the upper bound from the oblique parameters, and we impose all the constraints as discussed above.}
%%%

In Fig.~\ref{fig:ms-damu}, we show the scattering allowed plots in terms of muon $g-2$ and $m_S$.
It suggests that the typical value of muon $g-2$ is of the order $10^{-12}$ that is smaller than the experimental value by three order magnitude. 
%%%

In Fig.~\ref{fig:yukawa}, we demonstrate the couplings of $\kappa_1-\kappa_2$ in the left-figure, and $g_{L_1}-g_{L_2}$ in the right-figure. The left one implies $2\lesssim\kappa_1$ requires rather large coupling, whereas $\kappa_2$ is of the order 0.001, and  each of them has a weak correlation. 
While the right one suggests both of couplings run $-2.5\sim -0.3$ with degeneracy to some extent.

\section{Conclusion}
{We have constructed radiative neutrino mass model based on a gauged lepton flavor symmetry $U(1)_L$.
The condition to cancel gauge anomalies is discussed by introducing some exotic leptons with general $U(1)_L$ charge.  
Then we discuss phenomenology of the model by fixing the charge assignment.

The neutrino mass matrix can be induced at two-loop level where the exotic leptons and charged scalar bosons propagate inside the loop diagram.
{On the other hand the first and second charged-leptons of SM are induced at one-loop level.}
Due to the feature of flavor dependent symmetry, we have predicted one massless active neutrino and a bosonic dark matter candidate from inert doublet. 
Calculating the relic density, we have found that observed value can be obtained via Higgs portal interaction with mass range of dark matter at $56$ GeV$\sim$ $62.5$ GeV.  
In addition, we have also discussed lepton flavor violation and muon $g-2$ in the model.}

Then we have done the global numerical analysis to satisfy all the constraints such as charged-lepton masses, neutrino mass differences its mixing, LFVs, and oblique parameters, within the range of DM mass.
Then we have found  the typical value of muon $g-2$ is of the order $10^{-12}$ that is smaller than the experimental value by three order magnitude. Also we have shown the typical Yukawa couplings of $\kappa_{1,2}$ and $g_{L_{1,2}}$, and found typical ranges and their correlations.

%{The resulting $\Delta a_\mu$ has finally been found to be $H_{2\alpha}$ term dominant, and we have shown the allowed range in fig.~\ref{fig:ms-h22} to satisfy the observed $\Delta a_\mu=(26.1\pm8.0)\times10^{-10}$ in term of $m_s$ and $H_{22}$. The figure suggests that ${\cal O}(1.7)\lesssim H_{22}$ and $m_s\lesssim {\cal O}(300)$ GeV are favored, and the higher $m_s$ requires the larger $H_{22}$.}
%We have also shown the allowed region in terms of $M_{Z'}$ and $g_{Z'}$ to satisfy all the data that we have discussed before, and found $M_{Z'}\lesssim 0.3$ GeV.
%\newpage
%%%%%%%%%%%%%%%%%%%%%%%%%%%%%%%%%%%
%\hspace{0.2cm} {\bf Acknowledgments}
%\section*{Acknowledgments}:
%\vspace{0.5cm}
\section*{Acknowledgments}
\vspace{0.5cm}
H. O. is sincerely grateful for the KIAS member and all around.
%%%%%%%%%%%%%%%%%%%%%%%%%%%%%%%%%%%
%%%%%%%%%%%%%%%%%%%%%%%%%%%%%%%%%%%

\end{document}